\documentclass{aip-cp}

\usepackage[numbers]{natbib}
\usepackage{rotating}
\usepackage{graphicx}

\begin{document}

\title{Partners of the $X(3872)$ and HQSS breaking}

\author[aff1]{D.R. Entem\corref{cor1}}
\author[aff2]{P.G. Ortega}
\author[aff1]{F. Fern\'andez}

\affil[aff1]{Grupo de F\'isica Nuclear and IUFFyM, Universidad de Salamanca, E-37008 Salamanca, Spain.}
\affil[aff2]{CERN (European Organization for Nuclear Research), CH-1211 Geneva, Switzerland.}
\corresp[cor1]{Corresponding author: entem@usal.es}

\maketitle

\begin{abstract}
  Since the discovery of the $X(3872)$ the study of heavy meson molecules has been the subject of many investigations. On the experimental side
  different experiments have looked for its spin partners and the bottom analogs. On the theoretical side different approaches have been used to
  understand this state. Some of them are EFT that impose HQSS and so they make predictions for the partners of the $X(3872)$, suggesting the
  existence of a $J^{PC}=2^{++}$ partner in the charm sector or $J^{PC}=1^{++}$ or $2^{++}$ analogs in the bottom.

  In our work, in order to understand the $X(3872)$, we use a Chiral quark model in which, due to the proximity to the $DD^*$ threshold, we include
  $c\bar c$ states coupled to $DD^*$ molecular components. In this coupled channel model the relative position of the bare $c\bar c$ states with
  two meson thresholds are very important. We have looked for the $X(3872)$ partners and we don't find a bound state in the $D^*D^*$ $J^{PC}=2^{++}$.
  In the bottom sector we find the opposite situation where the $B^*B^*$ with $J^{PC}=2^{++}$ is bounded while the $J^{PC}=1^{++}$ is not bounded.
  These results shows how the coupling with $c\bar c$ states can induced different results than those expected
  by HQSS. The reason is that this symmetry is worse in the open heavy meson sector than in the hidden
  heavy meson sector.
\end{abstract}

\section{INTRODUCTION}

The theoretical study of exotic states, not well accommodated in the naive quark model, has motivated
a great interest since the discovery of the $X(3872)$. This state was discovered by the Belle collaboration
in 2003~\cite{PhysRevLett.91.262001} and very soon after confirmed by the 
CDF~\cite{PhysRevLett.93.072001}, D0~\cite{PhysRevLett.93.162002} and BaBar~\cite{PhysRevD.71.071103} collaborations.
The state lies well below where the $\chi_{c1}(2P)$ state is expected by quark models and is very close to the
$DD^*$ threshold. However it also presents very peculiar decay properties, being the most relevant the decays
into $J/\Psi \pi\pi$ through a $\rho$ and $J/\Psi \pi\pi\pi$ through an $\omega$ which implies
some kind of isospin violation. These properties rule out completely a $c\bar c$ interpretation of the state.
However it can be easily understood in the molecular picture as a $DD^*$ molecule. Then the isospin violation can be obtained
due to the difference on the mass of different charged states of the charmed mesons without the need
of introducing isospin violating interactions. 
For these reasons it is now accepted as the best candidate to be a non $q\bar q$ state.

The first consequence of the molecular picture is the existence of other analog states. In fact, if you
assume heavy flavor symmetry (HFS), which is an approximate good symmetry of QCD and implies that
the interactions are the same when you change a $c$ quark (antiquark) by a $b$ quark (antiquark),
then a $J^{PC}=1^{++}$ analog in the bottomonium sector should exist due to the reduction of the
kinetic energy by the bigger mass of the bottom mesons. So you should expect a state with bigger
binding energy and probably smaller isospin breaking properties due to the small difference in the masses
of the bottom mesons.

Other approximate symmetries can be used to study the existence of other analogs like, for instance,
heavy quark spin symmetry (HQSS) which implies the independence of the interactions on the spin of
the heavy quark. This symmetry has been use by the authors of Reference~\cite{PhysRevD.86.056004} to study the analogs
in the charmonium sector with other quantum numbers. The strongest conclusion of this work is the
existence of a $J^{PC}=2^{++}$ $B^*B^*$ analog which has been called $X(4012)$ since the interaction
in this channel is the same as in the channel of the $X(3872)$. In order to get conclusions for other
quantum numbers further assumptions are needed. Assuming that the $X(3915)$ is a $0^{++}$ analog
a total of six molecular states are found in the charmonium sector.

Using similar ideas and HFS the authors of Reference~\cite{PhysRevD.88.054007} studied the bottomonium sector
where the $1^{++}$ analog should be found and extended the study to the isovector channels in both sectors.

Although the pure molecular picture is very popular one could expect that the molecular states could couple
to nearby $c\bar c$ states. We followed this investigation in the framework of the Chiral Quark Model (CQM)
in Reference~\cite{PhysRevD.81.054023}. In this work we found that the $J^{PC}=1^{++}$ $DD^*$ channel is unbounded if we do not
include the coupling to $c\bar c$ states and is bounded once the coupling to the $\chi_{c1}(2P)$ is included.
In a latter work~\cite{JPhysG.40.651073} we also include other channels like the $J/\Psi \omega$ finding it
irrelevant. The reason is that the coupling of $J/\Psi\omega$ and $c\bar c$ states is an OZI forbidden process and so very suppressed
and the coupling to $DD^*$ states goes through a rearrangement process which is also very suppressed. So the most
important components are the $DD^*$ and $\chi_{c1}(2P)$. 

The irrelevance of the $J/\Psi\omega$ channel has been confirmed recently in lattice simulations~\cite{PhysRevLett.111.192001}
and also the strong coupling between the $DD^*$ components and $c\bar c$. In this work only the bound $DD^*$ state
is found and the expected $\chi_{c1}(2P)$ state is not found. For this reason it has been interpreted as a $c\bar c$ 
largely dressed by the $DD^*$ components. This differs from our picture since for us the $X(3872)$ appears as an additional
state to the naive quark model expectations.

If one assumes that the coupling to $c\bar c$ states can be important one important question is if this coupling
can make results deviate from expectations from HQSS and HFS. This is what we analyze in the present work.

\section{THE CHIRAL QUARK MODEL}

The present work use the CQM which has been extensively used to study baryon and meson
phenomenology. The $SU(2)$ version was published in Reference~\cite{JPhysG.19.2013} and later on was
extended to the $SU(3)$ version and to the heavy quark sector in Reference~\cite{JPhysG.31.481}.
Here we only describe the main ingredients of the model.

QCD has Chiral Symmetry as an approximate good symmetry at the level of the Lagrangian. However
this symmetry is not realized in the hadron spectrum and is spontaneously broken
at the quantum level. This fact implies the existence of massless Goldstone bosons (the pions) with the quantum
numbers of the broken symmetry. However the symmetry is not exact and the Goldstone bosons acquire a
dynamical (small) mass which is the most satisfactory way to understand the smallness of the mass of the
pion meson. When we break the symmetry in the $SU(3)$ version then the kaons and the eta meson appears.
These new degrees of freedom generates interactions between quarks as one Goldstone boson exchanges. Multiboson
exchanges are not included but they are taken into account by the exchange of scalar bosons.
However Chiral Symmetry Breaking also generates a dynamical mass for quarks, the constituent quark mass.

Besides the Chiral Symmetry Breaking another crucial non-perturbative effect of QCD is confinement
which makes all hadrons to appear as color singlets. We include confinement in a phenomenological way
as a color linear screened interaction between quarks.

QCD perturbative effects are included by the interaction induced by one gluon exchange which is of
special relevance on the description of heavy quark systems.

All the details of the interactions used can be found in References~\cite{JPhysG.31.481} and~\cite{PhysRevD.78.114033}
where the parameters used can be found.

\section{THE TWO MESON INTERACTION}

Once we have the interaction between the constituents of mesons we can study the interaction
between mesons generated by them. We obtain this interactions using the Resonating Group Method (RGM)
considering that we know the internal wave functions of the interacting mesons. In our calculation
these wave functions are obtained solving the two body ($q\bar q$) system.

In the systems considered in this work no antisymmetry effects between quarks in different mesons are
present. So the interaction is given by the so called RGM direct kernel
\begin{eqnarray}
    ^{RGM}V_D(\vec{P}',\vec{P}) &=& 
    \sum_{i\in A,j\in B} \int d\vec{p}_{\xi '_A}d\vec{p}_{\xi '_B}d\vec{p}_{\xi_A} d\vec{p}_{\xi_B} 
    \phi ^*_A (\vec{p}_{\xi '_A}) \phi ^*_B (\vec{p}_{\xi '_B}) 
    V_{ij}(\vec{P}',\vec{P}) \phi _A (\vec{p}_{\xi _A}) \phi _B (\vec{p}_{\xi _B})
\end{eqnarray}
However quarks or antiquarks can be exchanged between different mesons coupling different meson states, like,
for instance, the $DD^*\to J/\Psi\omega$. This kind of processes are rearrangement diagrams that are suppressed
by the meson wave functions overlaps.

\section{COUPLING MOLECULAR AND $Q\bar Q$ COMPONENTS}

As mentioned in the introduction the closest $Q\bar Q$ states to the two meson thresholds can coupled with these
states. A measured of this coupling is given by the widths of the $Q\bar Q$ states decaying to open heavy quark mesons,
which is of the order or tens or hundreds of MeV. 

In order to study this effect we use the $^3P_0$ model to coupled two quark and four quark sectors. It is important to notice
that the model only introduces a coupling that can be fitted to any decay and then all the others are given by
the meson wave functions and quark symmetries. In order to get an overall good description of two meson decays we
performed an analysis of strong decays in different sectors and we allowed the coupling to logarithmically run
with the scale of the system given by the reduced mass of the two quarks of the decaying meson. This was done
in Reference~\cite{Segovia2012322} where predictions in sectors where the coupling was not fitted were given
with very good agreement with the experimental data.

Then we take the hadronic wave function given by
\begin{equation}
  | \Psi \rangle = \sum_\alpha c_\alpha |\psi\rangle 
   + \sum_\beta \chi_\beta(P)|\phi_{M1}\phi_{M2}\beta\rangle
\end{equation}
where $|\psi\rangle$ are the hidden heavy mesons and $|\phi_{M1}\phi_{M2}\beta\rangle$ are the two
meson states with $\beta$ quantum numbers.

The coupling with $Q\bar Q$ states induced an effective energy dependent potential between the two mesons
given by
\begin{equation}
  V_{\beta'\beta}^{eff}(P',P) = \sum_\alpha \frac{h_{\beta'\alpha}(P')h_{\alpha\beta}(P)}{E-M_\alpha}
\end{equation}
We solve the coupled channel problem following Reference~\cite{EPJA44.93}. All the details can be found in
Reference~\cite{JPhysG.40.651073}.

\section{HEAVY QUARK SPIN SYMMETRY}

Since we want to analyze the discrepancies with HQSS expectations induced in our model by the coupling to $Q\bar Q$ states
is important to check if we don't have additional HQSS breaking effects. In our model HQSS breaking is present
since we take finite heavy quark masses, however one should expect small effects due to the big values of the
heavy quark masses.

\begin{figure}
\centerline{%
\includegraphics[width=.47\textwidth]{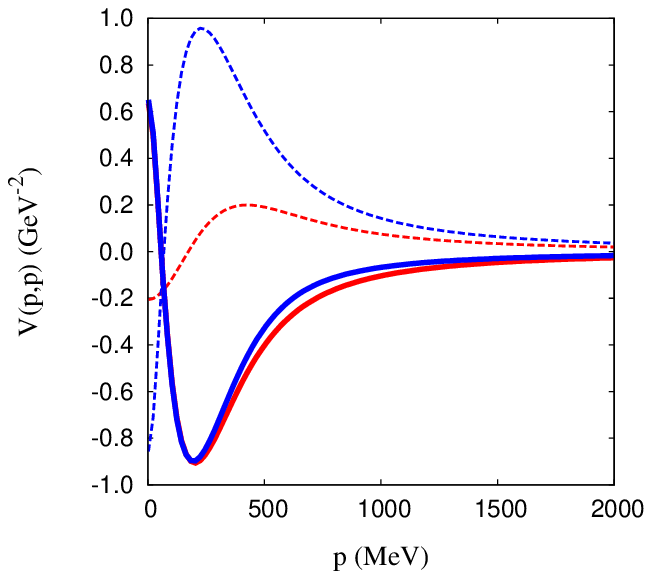}
\includegraphics[width=.47\textwidth]{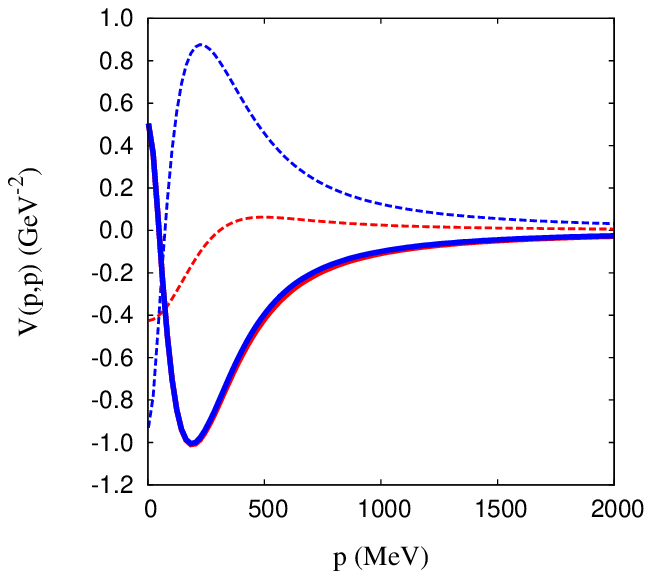}}
\caption{Diagonal matrix elements of the two meson interaction in momentum space for the 
  $D^{(*)}D^{(*)}$ sector (left panel)
  and $B^{(*)}B^{(*)}$ sector (right panel).
  The dashed blue line gives the $D^*D^*(0^{++})$ matrix element,
  the dashed red line the $DD(0^{++})$, the solid blue line the right hand side of Equation~\ref{Ec4} and the
solid red line left hand side of the same Equation.}
\label{fig1}
\end{figure}

\begin{figure}
\centerline{%
\includegraphics[width=.47\textwidth]{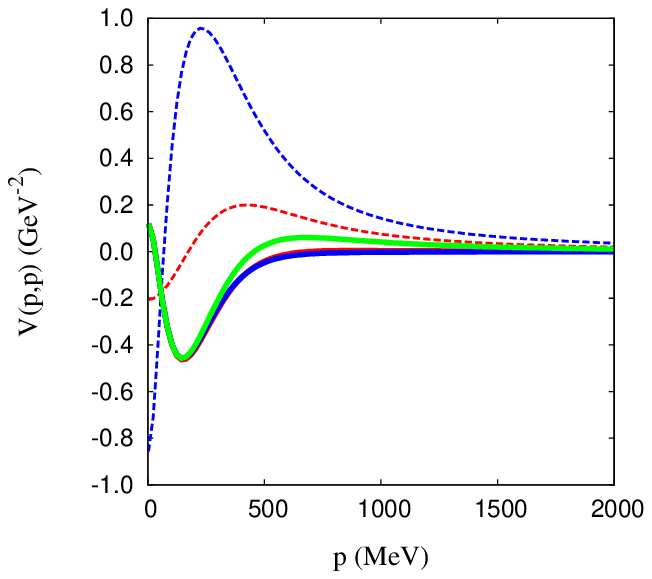}
\includegraphics[width=.47\textwidth]{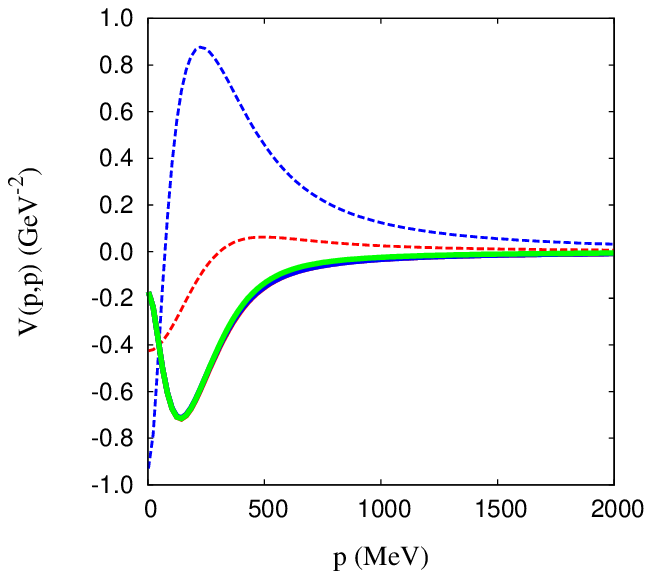}}
\caption{Diagonal matrix elements of the two meson interaction in momentum space for the 
  $D^{(*)}D^{(*)}$ sector (left panel)
  and $B^{(*)}B^{(*)}$ sector (right panel).
  The dashed blue line gives the $D^*D^*(0^{++})$ matrix element,
  the dashed red line the $DD(0^{++})$, the solid blue line the right hand side of Equation~\ref{Ec5}, the
  solid red line left hand side of the same Equation and the solid green line the right hand side of Equation~\ref{Ec6}.}
\label{fig2}
\end{figure}

\begin{figure}
\centerline{%
\includegraphics[width=.47\textwidth]{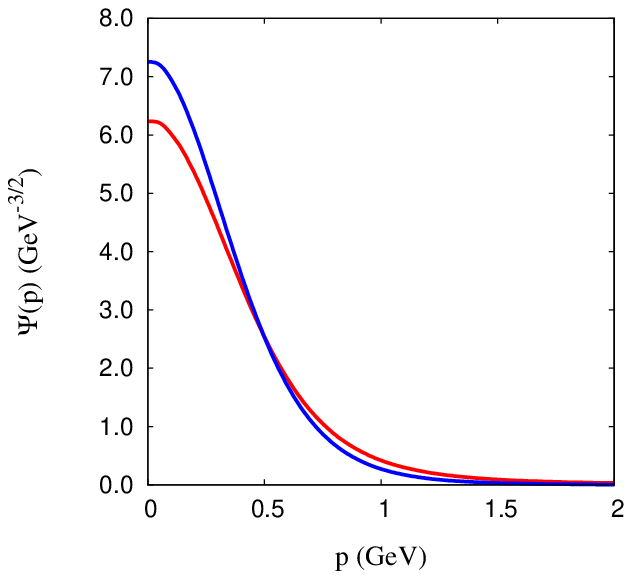}
\includegraphics[width=.47\textwidth]{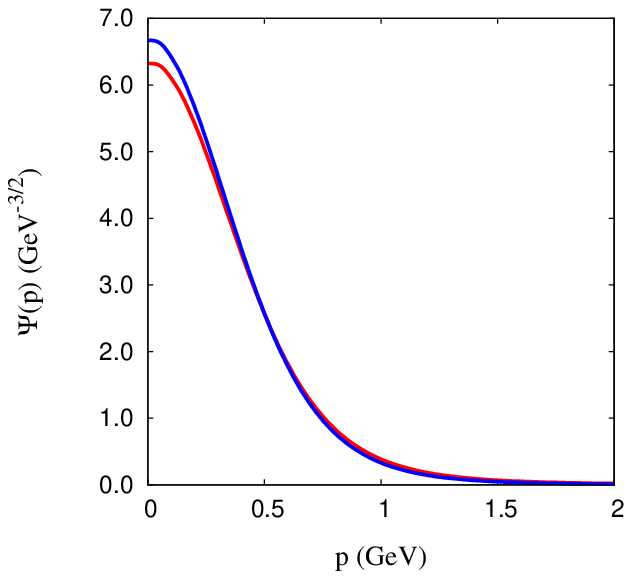}}
\caption{Wave function for the pseudoscalar (red) and vector (blue) heavy mesons. The left
  panel corresponds to charm mesons ($D^{(*)}$) and the right
  panel to bottom mesons ($B^{(*)}$).}
\label{fig3}
\end{figure}

For $S$-wave two meson states it is easy to find the following relations between matrix elements
\begin{eqnarray}
\frac{2}{\sqrt 3} \langle D^*D^* (0^{++}) | H |DD (0^{++}) \rangle &=&
\langle DD (0^{++}) | H |DD (0^{++}) \rangle
-\langle D^*D^* (0^{++}) | H |D^*D^* (0^{++}) \rangle
\label{Ec4}
\\
\langle DD^* (1^{++}) | H |DD^* (1^{++}) \rangle &=&
\langle D^*D^* (2^{++}) | H |D^*D^* (2^{++}) \rangle
\label{Ec5}
\\ &=&
\frac 3 2
\bigg[
\langle DD (0^{++}) | H |DD (0^{++}) \rangle
 -\frac 1 3 \langle D^*D^* (0^{++}) | H |D^*D^* (0^{++}) \rangle \bigg]
\label{Ec6}
\\
2 \langle DD^* (1^{+-}) | H |DD^* (1^{+-}) \rangle &=&
\langle DD (0^{++}) | H |DD (0^{++}) \rangle
+\langle D^*D^* (0^{++}) | H |D^*D^* (0^{++}) \rangle
\end{eqnarray}
which are given just by recoupling coefficients. These relations are checked in Figures~\ref{fig1} and~\ref{fig2}. For
exact HQSS all the solid lines should be the same, so we see just small HQSS breaking effects in the two meson interaction.
The small breaking effects are induced by the small difference in the wave functions of the pseudoscalar and vector heavy mesons which are
shown in Figure~\ref{fig3} and for exact HQSS should be the same.

\section{RESULTS}

\begin{figure}
\centerline{%
\includegraphics[width=.47\textwidth]{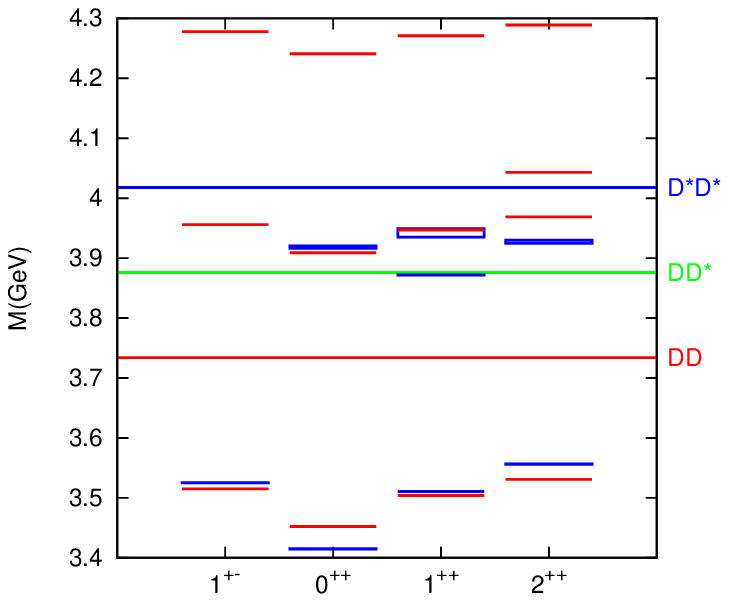}
\includegraphics[width=.47\textwidth]{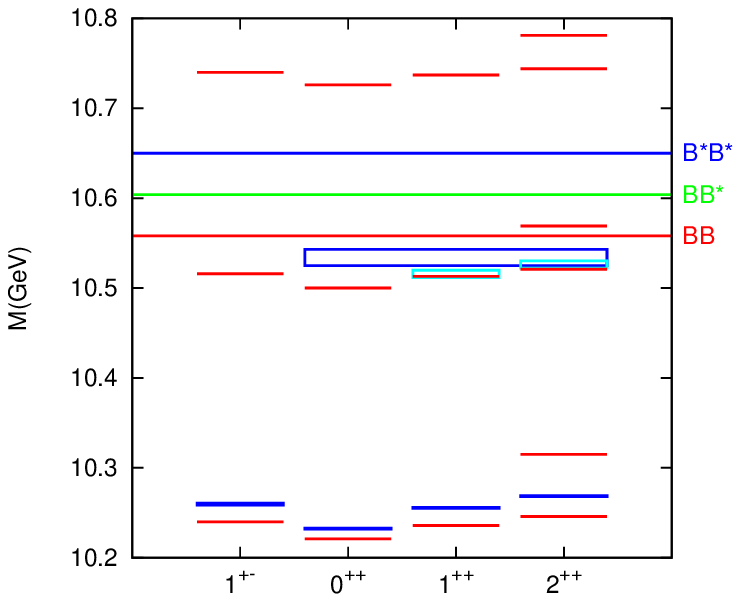}}
\caption{Charm (left panel) and bottom (right panel) spectrum. Blue boxes
  shows the states in the Particle Data Group~\cite{PDG}. See comments in the text for
  the PDG states selected. 
  We also have included in light blue the candidates for the $\chi_{b1}(2P)$ and $\chi_{b2}(2P)$ 
  measured recently by LHCb~\cite{JHEP10.088}.
  The states
in red are the pure $Q\bar Q$ states predicted by the model.}
\label{fig4}
\end{figure}

As shown in the previous section the model preserves HQSS in a good approximation although small breaking
effects are present due to the use of finite heavy quark masses. However it is interesting to
notice that the HQSS breaking in the masses of open charm and bottom mesons is bigger than in the
hidden charm and hidden bottom sectors. For this reason the relations between the two meson
thresholds and the $Q\bar Q$ states change when we change the $J^{PC}$ quantum numbers. This can be seen
in Figure~\ref{fig4} where we present the predictions of the pure $Q\bar Q$ states of the model in red
compared with the two meson thresholds and the states of the Particle Data Group (PDG) from Reference~\cite{PDG} in blue.
For the PDG states we only include states with well known quantum numbers with some exceptions. In the $1^{+-}$ sector
although the $I^G$ quantum numbers has not been measured it is a well accepted candidate for the $h_c(1P)$ state
as shown by the agreement with our quark model result. For the $X(3940)$ the put in the $1^{++}$ sector since it
has been seen in $DD^*$ decays and so it is a good candidate for the $\chi_{c1}(2P)$ state.
In the bottomonium sector the same applies to the $h_b(1P)$ and $h_b(2P)$ (included in the updated version).
For the $\chi_b(3P)$ only $C=+$ has been measured and we have included as a state covering the three possible
assignments. We also have included in light blue the candidates for the $\chi_{b1}(2P)$ and $\chi_{b2}(2P)$ 
measured recently by LHCb~\cite{JHEP10.088}.

It is important to consider the following aspects. In the charmonium sector the relevant thresholds, in which an
$S$-wave two meson state is possible,
are $DD^*$ and $D^*D^*$ in $1^{+-}$, $DD$ and $D^*D^*$ in $0^{++}$, $DD^*$ in $1^{++}$ and $D^*D^*$ in $2^{++}$.
The same occurs in the bottomonium sector changing the $D^{(*)}$ mesons by $B^{(*)}$ mesons.
Then, a $Q\bar Q$ state above threshold gives attraction and below gives repulsion. The strength is inversely proportional
to the difference between the threshold position and the mass of the bare $Q\bar Q$ state.

Our prescription in the present work is to include the closer state above and below the relevant threshold and
we only include two meson states where an $S$-wave is present. The partial waves included for two meson states are
\begin{enumerate}
\item In the $0^{++}$ sector we include $DD$ ($BB$) and $D^*D^*$ ($B^*B^*$) $^1S_0$ waves
\item In the $1^{++}$ sector we include $DD^*$ ($BB^*$) $^3S_1$ and $^3D_1$ waves
\item In the $1^{+-}$ sector we include $DD^*$ ($BB^*$) and $D^*D^*$ ($B^*B^*$) $^3S_1$ and $^3D_1$ waves
\item In the $2^{++}$ sector we include $D^*D^*$ ($B^*B^*$) $^5S_2$, $^1D_2$ and $^5D_2$ waves
\end{enumerate}

The results are summarized in Table~\ref{tab1} where we only quote additional states to the dressed $Q\bar Q$ states that
we find as resonances.

\begin{table}
\caption{Additional states to the dressed $Q\bar Q$ states given by the model.
  The states predicted using HFS and HQSS symmetries given in References~\cite{PhysRevD.86.056004,PhysRevD.88.054007}
are given for comparison.
\label{tab1}}
\begin{tabular}{lcccc}
\hline
Charmonium              & $1^{+-}$ & $0^{++}$ & $1^{++}$ & $2^{++}$ \\
\hline
Reference~\cite{PhysRevD.86.056004}& 3815/3955 & 3710/Input & Input    &  4012    \\
Reference~\cite{PhysRevD.88.054007}&           &            & Input    &  4012     \\
This work         &           &            & $X(3872)$&      \\
\hline 
\hline
Bottomonium       & $1^{+-}$  & $0^{++}$   & $1^{++}$ & $2^{++}$ \\
\hline
Reference~\cite{PhysRevD.88.054007}&           &            &  10580   &  10600    \\
This work         &           &  10621     &    ?     &  10648   \\
\hline
\end{tabular}
\end{table}

Let's consider first the $1^{++}$ $DD^*$ and $2^{++}$ $D^*D^*$ channels where HQSS tells us that the interaction is
the same. In the charm sector we see that in the $1^{++}$ we get more attraction than repulsion while in the
$2^{++}$ we get repulsion (the state above threshold is an $F$-wave $c\bar c$ state that is weakly coupled to the
two meson sector). As shown in Reference~\cite{PhysRevD.81.054023} the two meson interaction is not enough to bind the
system, however the coupling with the $\chi_{c1}(2P)$ state gives the additional attraction and we get a new state
that we assigned to the $X(3872)$. In the $2^{++}$ sector we don't get this additional attraction and for this
reason no new state is found. However the dressed $c\bar c(^3F_2)$ state gets dressed and a state close to the predicted
$X(4012)$ by HQSS is found but not present as an additional state. For this reason although an state with similar mass
is found we expect different decays properties.

In the bottomonium we have the opposite situation. Now the two meson interaction is enough to bind the system. 
In the $1^{++}$ we expect more repulsion
from the state below the $BB^*$ threshold. In Table~\ref{tab1} we don't give an answer to the existence or not
existence of this state since we can find a very shallow bound state or no state when we vary the value of the
strength parameter of the $^3P_0$ model within its uncertainties.
In the $2^{++}$ we get 
similar repulsion and attraction from the
states below and above threshold and here we find an additional state. This situation differs completely from the
model independent expectations of HQSS and HFS without coupling to $Q\bar Q$ states.

For the other quantum numbers HQSS need some assumptions and the final result depends on these assumptions. In Table~\ref{tab1}
we compare our results in the $0^{++}$ and $1^{+-}$ sectors with the conclusions from 
References~\cite{PhysRevD.86.056004,PhysRevD.88.054007}
where a better agreement is found with Reference~\cite{PhysRevD.88.054007}.

As a summary, in this work we only include the closest $Q\bar Q$ state above and below the relevant thresholds, however a
more complete description including all the closest states is necessary to give definite conclusions. This will
be presented in a forthcoming paper. However the main conclusion of this work is that the coupling between
two meson states and $Q\bar Q$ states can modified HQSS and HFS conclusions in the pure molecular picture, and
we think that this problem should be address in models at hadron level.

\section{ACKNOWLEDGMENTS}
This work has been partially funded by Ministerio de Ciencia y Tecnolog\'\i a
under Contract no. FPA2013-47443-C2-2-P, 
by the Spanish Excellence Network on Hadronic Physics
FIS2014-57026-REDT
and by the Spanish Ingenio-Consolider 2010
Program CPAN (CSD2007-00042). 
P.G. Ortega
acknowledges the financial support from the European Union's Marie Curie COFUND
grant (PCOFUND-GA-2011-291783).


\nocite{*}
\bibliographystyle{aipnum-cp}%
\bibliography{entem}%

\end{document}